\documentclass[prd,superscriptaddress,amsfonts,amssymb,amsmath,showpacs,twocolumn]{revtex4-2}
\usepackage{bm}
\usepackage{amsfonts}
\usepackage{latexsym}
\usepackage[latin1]{inputenc}
\usepackage{graphicx}
\usepackage{amsmath}
\usepackage{palatino}
\usepackage{ragged2e}
\usepackage{mathpazo}
\usepackage{textcomp}
\usepackage{float}
\usepackage{booktabs}
\usepackage{dcolumn}
\usepackage{multirow}
\usepackage{hyperref}
\hypersetup{colorlinks,citecolor=blue}
\usepackage{amsmath}
\usepackage{xcolor}
\usepackage{orcidlink}
\usepackage[caption=false]{subfig}
\usepackage{commath}
\def\jnl@style{\it}
\def\aaref@jnl#1{{\jnl@style#1}}

\def\aaref@jnl#1{{\jnl@style#1}}

\def\aj{\aaref@jnl{AJ}}                   
\def\apj{\aaref@jnl{ApJ}}                 
\def\apjl{\aaref@jnl{ApJ}}                
\def\apjs{\aaref@jnl{ApJS}}               
\def\apss{\aaref@jnl{Ap\&SS}}             
\def\aap{\aaref@jnl{A\&A}}                
\def\aapr{\aaref@jnl{A\&A~Rev.}}          
\def\aaps{\aaref@jnl{A\&AS}}              
\def\mnras{\aaref@jnl{Mon.~Not.~Roy.~Astron.~Soc.}}             
\def\prd{\aaref@jnl{Phys.~Rev.~D}}        
\def\prc{\aaref@jnl{Phys.~Rev.~C}}  
\def\prl{\aaref@jnl{Phys.~Rev.~Lett.}}    
\def\qjras{\aaref@jnl{QJRAS}}             
\def\skytel{\aaref@jnl{S\&T}}             
\def\ssr{\aaref@jnl{Space~Sci.~Rev.}}     
\def\zap{\aaref@jnl{ZAp}}                 
\def\nat{\aaref@jnl{Nature}}              
\def\aplett{\aaref@jnl{Astrophys.~Lett.}} 
\def\apspr{\aaref@jnl{Astrophys.~Space~Phys.~Res.}} 
\def\physrep{\aaref@jnl{Phys.~Rep.}}      
\def\physscr{\aaref@jnl{Phys.~Scr}}       
\def\commat{\aaref@jnl{Comm.~Math.~Phys.}}              
\def\science{\aaref@jnl{Science}}               
\def\cqg{\aaref@jnl{Classical Quant.~Grav.}}            
\def\jpcs{\aaref@jnl{JPCS}}                                     
\def\ijmpd{\aaref@jnl{Int.~J.~Mod.~Phys.~D}}                    
\def\grg{\aaref@jnl{Gen.~Relat.~Gravit.}}               
\def\rpp{\aaref@jnl{Rep.~Prog.~Phys.}}          
\def\npa{\aaref@jnl{Nucl.~Phys.~A}}        
\def\lrr{\aaref@jnl{Living Rev.~Rel.}}                   
\def\jcap{\aaref@jnl{J.~Cosmology Astropart.~Phys.}}    
\def\rmp{\aaref@jnl{Rev.~Mod.~Phys.}}   
\def\epjc{\aaref@jnl{Eur.~Phys.~J.~C}}

\allowdisplaybreaks[1]

\begin{document}

\color{black}       

\title{Cosmic acceleration with bulk viscosity in an anisotropic $f(R,L_m)$ background}

\author{Raja Solanki\orcidlink{0000-0001-8849-7688}}
\email{rajasolanki8268@gmail.com}
\affiliation{Department of Mathematics, Birla Institute of Technology and
Science-Pilani,\\ Hyderabad Campus, Hyderabad-500078, India.}
\author{Bina Patel\orcidlink{0000-0002-3079-0061}}
\email{binapatel.maths@charusat.ac.in}
\affiliation{Department of Mathematical Sciences, P.D. Patel Institute of Applied Sciences, Charotar University of Science and Technology (CHARUSAT), Changa,
Anand 388421, Gujarat, India}
\author{Lakhan V. Jaybhaye\orcidlink{0000-0003-1497-276X}}
\email{lakhanjaybhaye@gmail.com}
\affiliation{Department of Mathematics, Birla Institute of Technology and
Science-Pilani,\\ Hyderabad Campus, Hyderabad-500078, India.}

\author{P.K. Sahoo\orcidlink{0000-0003-2130-8832}}
\email{pksahoo@hyderabad.bits-pilani.ac.in}
\affiliation{Department of Mathematics, Birla Institute of Technology and
Science-Pilani,\\ Hyderabad Campus, Hyderabad-500078, India.}

\date{\today}

\begin{abstract}
In this article, we investigate the observed cosmic acceleration in the framework of a  cosmological $f(R,L_m)$ model dominated by bulk viscous matter in an anisotropic background. We consider the LRS Bianchi type I metric and derive the Friedmann equations that drive the gravitational interactions in $f(R,L_m)$ gravity. Further, we assume the functional form $f(R,L_m)=\frac{R}{2}+L_m^\alpha $, where $\alpha$ is a free model parameter, and then find the exact solutions of fields equations corresponding to our viscous matter dominated model. We incorporate the updated H(z) data and the Pantheon data to acquire the best-fit values of parameters of our model by utilizing the $\chi^2$ minimization technique along with the Markov Chain Monte Carlo (MCMC) random sampling method. Further, we present the behavior of physical parameters that describe the universe's evolution phase, such as density, effective pressure and EoS parameters, skewness parameter, and the statefinder diagnostic parameters. We found that the energy density indicates expected positive behavior, whereas the negative behavior of bulk viscous pressure contributes to the universe's expansion. The effective EoS parameter favors the accelerating phase of the universe's expansion. Moreover, the skewness parameter shows the anisotropic nature of spacetime during the entire evolution phase of the universe. Finally, from statefinder diagnostic test, we found that our cosmological $f(R,L_m)$ model lies in the quintessence region, and it behaves like a de-Sitter universe in the far future. We analyze different energy conditions in order to test the consistency of the obtained solution. We found that all energy conditions except SEC show positive behavior, while the violation of SEC favors the recently observed acceleration with the transition from decelerated to an accelerated epoch of the universe's expansion in the recent past.

\end{abstract}

\maketitle

\section{Introduction}

\justify The universe in the beginning and future may have anisotropies despite the widely accepted assumption that the current universe is homogeneous and isotropic. In 1992, Cosmic Background Explorer \cite{COBE} revealed that the cosmic microwave background possesses a small amount of anisotropy at cosmological scales. The observational evidence, such as Wilkinson Microwave Anisotropy Probe experiments \cite{WMAP} and Cosmic Background Imager \cite{CBI}, strongly support this anisotropic nature of spacetime geometry. In 1998, the Supernovae Cosmology Project led by Perlmutter and Riess confirmed the accelerating behavior of the universe's expansion \cite{Riess,Perl}. Moreover, recent advancements indicate the variations in the strengths of microwaves received from different directions, i.e., the universe's expansion is anisotropic \cite{KMIG}. One can efficiently describe the homogeneous and anisotropic nature of spacetime using Bianchi-type cosmology. Several models based on Bianchi cosmology have appeared in the literature \cite{Bnc-1,Bnc-2,Bnc-4,Bnc-5,Bnc-5c,Bnc-5d,Bnc-5e}. In the present work, we incorporate locally rotationally symmetric (LRS) Bianchi type-I metric that is near to the FLRW metric and can be considered as a generalization of the isotropic one.

\justify It is well known that the modified theories of gravity are capable of bypassing the undetected dark energy issue. The cosmological models based on the modifications of the Einstein action of general relativity (GR) that are obtained by introducing the function $f(R)$ of the Ricci curvature $R$, first appeared in \cite{H.A.,R.K.,H.K.}. This cosmological model could describe the observed cosmic expansion without incorporating any dark energy component \cite{Carr,Cap}. Observational characteristics of $f(R)$ gravity models, along with the constraints acquired by equivalence principle and solar system tests, are investigated in the references \cite{Shin,Sal,Alex,PP1,PP2}. The viable $f(R)$ gravity models in the perspective of solar system tests do exist \cite{Noj,V.F.,L.A.}. Odintsov et al. have studied energy conditions and the $H_0$ tension in the context of $f(R)$ gravity models \cite{Odi-1,Odi-2}. One can follow references \cite{Noj-2,Noj-3,JS,Noj-4,Odi-3,Odi-4,AP} to review some viable $f(R)$ gravity models.

\justify Bertolami et al. \cite{O.B.} introduced a generalization of the $f(R)$ modified gravity by assuming an explicit coupling between the generic $f(R)$ function of the Ricci curvature scalar $R$ and the Lagrangian density of matter $L_m$, in the Einstein-Hilbert action. T. Harko and F. S. N Lobo further extended this model to the case of arbitrary geometry-matter couplings \cite{THK}. The cosmological models incorporating the non-minimal curvature and matter couplings present some interesting applications in cosmology and astrophysics \cite{THK-2,THK-3,V.F.-2}. Further, T. Harko and F. S. N Lobo proposed \cite{THK-6} $f(R,L_m)$ gravity, which is the set of all curvature-matter coupling theories. In this modified gravity, the energy-momentum tensor has the non-vanishing covariant divergence, an extra force orthogonal to four velocities arises, and the motion of the test particle is non-geodesic. Moreover, the $f(R,L_m)$ modified theory of gravity does not satisfy the equivalence principle, and it is constrained by the solar system tests \cite{FR,JP}. Nowadays, plenty of interesting cosmological implications of $f(R,L_m)$ gravity theory started appearing in the literature; for instance, check references \cite{GM,RV-1,RV-2,Jay,THK-7,THK-8,Kav}. 

\justify In this manuscript, we investigate the bulk viscous matter dominated $f(R,L_m)$ cosmological model with the anisotropic background. The significance of viscosity coefficients in cosmological modeling has a long history. From the hydrodynamics perspective, whenever a system loses its thermal equilibrium state, an effective pressure is generated to retrieve its thermal stability. Bulk viscosity in a matter content of the cosmos can be viewed as a manifestation of such an effective pressure. We aim to investigate the impact on the evolution phase of the universe when we consider the effect of bulk viscosity coefficient $\zeta$ in the usual cosmic pressure. We assume that the coefficient of viscosity $\zeta$ obeys a scaling law, and that reduces the Einstein case to a form proportional to the Hubble parameter \cite{IB-0}. This scaling law considered in our study is quite useful. One can follow references to review some attractive cosmological models incorporating the viscosity in cosmic fluid \cite{IB-2,IB-3,JM,AVS,MAT,VV5,VV6}.

The present work is organized as follows. In Sec \ref{sec2}, we present the fundamental formulations of $f(R,L_m)$ gravity. In Sec \ref{sec3}, we derive the field equations corresponding to the LRS Bianchi type I metric. Further, we consider a cosmological $f(R,L_m)$ model and then obtain the Hubble parameter value in terms of cosmic redshift. In Sec \ref{sec4}, we utilize the H(z) and Pantheon observational data to estimate the model and bulk viscous parameter values. In the next section Sec \ref{sec5}, we present the physical behavior of cosmological parameters and statefinder diagnostic corresponding to the values of parameters obtained by the H(z), Pantheon, and H(z)+Pantheon data. In addition, in Sec \ref{sec6}, we investigate the different energy conditions. Finally, in Sec \ref{sec7}, we discuss and conclude our results.

\section{ $f(R,L_m)$ Gravity Theory}\label{sec2}

\justify

The action ruling the gravitational interactions in $f(R,L_m)$ gravity is given as

\begin{equation}\label{1a}
S= \int{f(R,L_m)\sqrt{-g}d^4x}.
\end{equation}
Here $f(R,L_m)$ is the generic function of the Ricci curvature $R$ and the matter Lagrangian $L_m$.

\justify The Ricci curvature term $R$ can be obtained by the following contraction,
\begin{equation}\label{1b}
R= g^{\mu\nu} R_{\mu\nu},
\end{equation}
where $R_{\mu\nu}$ is the Ricci tensor and it is defined as

\begin{equation}\label{1c}
R_{\mu\nu}= \partial_\lambda \Gamma^\lambda_{\mu\nu} - \partial_\mu \Gamma^\lambda_{\lambda\nu} + \Gamma^\lambda_{\mu\nu} \Gamma^\sigma_{\sigma\lambda} - \Gamma^\lambda_{\nu\sigma} \Gamma^\sigma_{\mu\lambda},
\end{equation}
with $\Gamma^\alpha_{\beta\gamma}$ denoting the components of Levi-Civita connection.

\justify One can obtained the following field equation of $f(R,L_m)$ gravity by varying the action \eqref{1a} with respect to the metric $g_{\mu\nu}$,

\begin{equation}\label{1d}
f_R R_{\mu\nu} + (g_{\mu\nu} \square - \nabla_\mu \nabla_\nu)f_R - \frac{1}{2} (f-f_{L_m}L_m)g_{\mu\nu} = \frac{1}{2} f_{L_m} T_{\mu\nu},
\end{equation}

\justify where $f_R \equiv \frac{\partial f}{\partial R}$, $f_{L_m} \equiv \frac{\partial f}{\partial L_m}$, and $T_{\mu\nu}$ is the energy momentum tensor given by

\begin{equation}\label{1e}
T_{\mu\nu} = \frac{-2}{\sqrt{-g}} \frac{\delta(\sqrt{-g}L_m)}{\delta g^{\mu\nu}}.
\end{equation}

\justify One can acquired the following relation by contracting the field equation \eqref{1d}

\begin{equation}\label{1f}
R f_R + 3\square f_R - 2(f-f_{L_m}L_m) = \frac{1}{2} f_{L_m} T.
\end{equation}

\justify Here $R$, $T$, and $L_m$ being the Ricci curvature scalar, energy-momentum scalar, and the matter Lagrangian term respectively. Furthermore, $\square F = \frac{1}{\sqrt{-g}} \partial_\alpha (\sqrt{-g} g^{\alpha\beta} \partial_\beta F)$ for any scalar function $F$ .

\justify Moreover, by employing the covariant derivative in equation \eqref{1d} one can obtain the following relation 

\begin{equation}\label{1g}
\nabla^\mu T_{\mu\nu} = 2\nabla^\mu ln(f_{L_m}) \frac{\partial L_m}{\partial g^{\mu\nu}}.
\end{equation} 

\section{The Cosmological $f(R,L_m)$ Model with Anisotropic Background}\label{sec3}
\justify
Taking into account anomalies found in the observations of CMB, we consider following the LRS Bianchi type-I metric to describe the spatially homogeneous and anisotropic nature of the universe 
\begin{equation}\label{2a}
ds^2=-dt^2+A^2(t)dx^2+B^2(t)(dy^2+dz^2),
\end{equation}

\justify where $A(t)$ and $B(t)$ are time dependent metric potentials. In particular, one can recover the usual flat FLRW background for $A(t)=B(t)=a(t)$. Now, for the line element \eqref{2a}, we obtained the Ricci scalar as

\begin{equation}\label{2b}
R= 2 (\dot{H_x}+2\dot{H_y})+2({H_x}^2+3{H_y}^2)+4{H_x}{H_y}.
\end{equation}
\justify Here $H_x=\frac{\dot{A}}{A}$ and $H_y=H_z=\frac{\dot{B}}{B}$ represents directional Hubble parameters.

\justify The stress-energy tensor characterizing the universe filled with viscous content reads as
\begin{equation}\label{2c}
T_{\mu \nu}=(\rho+\bar{p})u_{\mu}u_{\nu} + \bar{p}g_{\mu \nu},
\end{equation}
where $\bar{p}=p-3 \zeta H$ denote the effective pressure of the cosmic fluid in the presence 
of viscosity coefficient $\zeta$ and $\rho$ is the usual matter energy density with four velocities components $u^\mu=(1,0,0,0)$. 

\justify The assumption of viscous effects in the cosmic fluid content can be viewed as an attempt to refine its description, relaxing its ideal properties and contributing negatively to the total pressure that can drive the aforementioned cosmic late-time acceleration. This can be checked in References \cite{odintsov/2020,fabris/2006,meng/2009}.

\justify The field equations governing the dynamics of the viscous matter dominated universe in $f(R,L_m)$ with anisotropic background reads as

\begin{widetext}
\begin{equation}\label{2d}
(2{H_x}{H_y} +{H_y}^2)f_R +\frac{1}{2}(f-Rf_R-L_{m}f_{L_m}) +\dot{f_R}(H_{x} +2H_{y}) =\frac{1}{2} f_{L_m}\rho ,
\end{equation}
\begin{equation}\label{2e}
(\dot{H_x}+{H_{x}}^2 +2{H_x}{H_y}){f_R}-\ddot{f_R}-\dot{f_R}({H_x}+{2H_y}) -\frac{1}{2}(f-{L_m}{f_{L_m}})  = \frac{1}{2} f_{L_m}\bar{p},
\end{equation}
\begin{equation}\label{2f}
(\dot{H_y} + 2{H_y}^2 +{H_x }{H_y}) {f_R} -\ddot{f_R}- \dot{f_R}(H_x +{2H_y}) -\frac{1}{2}(f-{L_m}{f_{L_{m}}}) =\frac{1}{2} f_{L_m}(\bar{p} +\delta \rho) .
\end{equation}
\end{widetext}

\justify Here $\delta$ is a measure of deviation from isotropy along $y$ and $z$ axes, called skewness parameter.

\justify In this work, we study the following $f(R,L_m)$ model in order to explore the cosmological implications of the $f(R,L_m)$ gravity \cite{LK,LB},

\begin{equation}\label{2g} 
f(R,L_m)=\frac{R}{2}+L_m^\alpha ,
\end{equation}

\justify where $\alpha$ is arbitrary parameter. The cosmological $f(R,L_m)$ model that we have  considered is motivated by the generic function $f(R,L_m) = f_1(R)+f_2(R) G(L_m)$ that represents arbitrary curvature-matter coupling \cite{LB}. In particular, for $\alpha=1$ one can obtain usual Friedmann equations of GR. 

\justify Now one can acquire the spatial volume element as
\begin{equation}\label{2h}
V=AB^2.
\end{equation}
\justify Then by using \eqref{2h}, the mean value of Hubble parameter can be calculated as
\begin{equation}\label{2i}
H=\frac{1}{3}\frac{\dot{V}}{V}= \frac{1}{3} (H_x+2H_y).
\end{equation}

\justify Further, we consider an extra ansatz to obtain the exact solution of the field equations. We consider a physical relation between the expansion scalar $\theta$ and shear scalar $\sigma$, specifically, $\theta^2 \propto \sigma^2$ . In particular, for $\frac{\sigma}{\theta}=constant$, one can retrieve the isotropy of Hubble expansion \cite{Ade}. This gives rise to a following relation
\begin{equation}\label{2j}
A=B^n,
\end{equation}
and the corresponding directional Hubble parameter can be obtained as
\begin{equation}\label{2k}
H_x=n H_y , n \neq 0.
\end{equation}
\justify For $n=1$, one can recover the usual flat FLRW cosmology.

\justify Now, for the specific function \eqref{2g} with $L_m=\rho$ \cite{HLR}, we obtained the following Friedmann equations for non-relativistic viscous matter dominated universe, by using equations \eqref{2d}-\eqref{2f} 

\begin{equation}\label{2l}
 (2n+1){H_y}^2 = (2\alpha-1){\rho}^{\alpha},
\end{equation}
\begin{equation}\label{2m}
 2\dot{H_y}+3{H_y}^2= (\alpha-1){\rho}^{\alpha}-{\alpha}{\rho^{\alpha-1}}\bar{p},
\end{equation}
\begin{equation}\label{2n}
(n+1)\dot{H_y}+(n^2+n+1){H_y}^2=(\alpha-1){\rho}^{\alpha}-\alpha{\rho^{\alpha-1}}(\bar{p}+\delta \rho).
\end{equation}

\justify By using equations \eqref{2l} and \eqref{2m}, we obtain following first-order equation
 
\begin{equation}\label{2o}
 \dot{H_y}+ \left[ \frac{5\alpha-2+2n(1-\alpha)}{2(2\alpha-1)} \right] H_y^2 = {\frac{\alpha \zeta}{2}}(n+2)\bigg(\frac{2n+1}{2\alpha-1}\bigg)^{\frac{\alpha-1}{\alpha}} {H_y}^{\frac{3\alpha-2}{\alpha}}.
 \end{equation}

\justify On integrating the above equation, we have 

\begin{widetext}
\begin{equation}\label{2p}
  H_y=\biggl\{ H_{y_0}^{\frac{2-\alpha}{\alpha}} (1+z)^{\frac{(2-\alpha)[5\alpha-2+2n(1-\alpha)]}{2\alpha(2\alpha-1)}} +\frac{\alpha \xi (n+2)(2\alpha-1)^{\frac{1}{\alpha}}(2n+1)^{\frac{\alpha-1}{\alpha}}}{5\alpha-2+2n(1-\alpha)} \big[ 1- (1+z)^{{\frac{(2-\alpha)[5\alpha-2+2n(1-\alpha)]}{2\alpha(2\alpha-1)}}}\big] \biggr\}^{\frac{\alpha}{2-\alpha}}.
\end{equation}
\end{widetext}

\justify Here $H_y(0)=H_{y_0}$. Now by using equations \eqref{2i} and \eqref{2k}, we obtained the average value of the Hubble parameter as
\begin{equation}\label{2q}
H= \big( \frac{n+2}{3} \big) H_y .  
\end{equation}
\justify By using equation \eqref{2q} in \eqref{2p}, we obtained the expression for averaged Hubble parameter in terms of redshift as

\begin{widetext}
\begin{equation}\label{2r}
  H(z)= \big( \frac{n+2}{3} \big) \biggl\{ \big( \frac{3H_0}{n+2} \big)^{\frac{2-\alpha}{\alpha}} (1+z)^{\frac{(2-\alpha)[5\alpha-2+2n(1-\alpha)]}{2\alpha(2\alpha-1)}} +\frac{\alpha \xi (n+2)(2\alpha-1)^{\frac{1}{\alpha}}(2n+1)^{\frac{\alpha-1}{\alpha}}}{5\alpha-2+2n(1-\alpha)} \big[ 1- (1+z)^{{\frac{(2-\alpha)[5\alpha-2+2n(1-\alpha)]}{2\alpha(2\alpha-1)}}}\big] \biggr\}^{\frac{\alpha}{2-\alpha}},
\end{equation}
\end{widetext}

\justify where $H(0)=H_0$ is the present value of the Hubble parameter.

\section{ Observational Constraints}\label{sec4}
\justify In this section, we are going to investigate the viability of our theoretical model by utilizing observational data sets. In this work, we incorporate updated H(z) data sets and the Pantheon supernovae observations. In order to perform statistical analysis, we use the $\chi^2$ minimization technique along with the Markov Chain Monte Carlo (MCMC) random sampling method in the emcee python library \cite{Mac}.

\subsubsection{H(z) datasets}
 It is well known that the expansion of the universe can be characterized by the Hubble parameter. For early passively evolving galaxies, one can recover the Hubble parameter values by measuring the $dt$ at a definite redshift value. In this work, we use $57$ points of $H(z)$ data in the redshift range $0.07 \leq z \leq 2.41$ \cite{GSS}. Moreover, there are two well-known approaches to extracting the values of the Hubble parameter, namely the differential age (DA) and line of sight BAO method.
 The set of $57$ H(z) data points can be checked in the reference \cite{RS,DDDD}. Now, the chi-square function corresponding to H(z) data points reads as

\begin{equation}\label{4a}
\chi _{H}^{2}(H_0,\alpha,\zeta,n)=\sum\limits_{k=1}^{57}
\frac{[H_{th}(z_{k},H_0,\alpha,\zeta,n)-H_{obs}(z_{k})]^{2}}{
\sigma _{H(z_{k})}^{2}},
\end{equation}

\justify where, the theoretically predicted value of the H(z) is denoted by $H_{th}$ whereas $H_{obs}$ represents the observed value of H(z) with standard error $\sigma_{H(z_{k})}$. 
The contour plot for the model parameters $\alpha$, $\zeta$, $n$, and $H_0$ corresponding to H(z) data sets at $1-\sigma$ and $2-\sigma$ confidence interval is presented in Fig \ref{f1}.

\begin{widetext}
\begin{center}
\begin{figure}[H]
\includegraphics[scale=0.8]{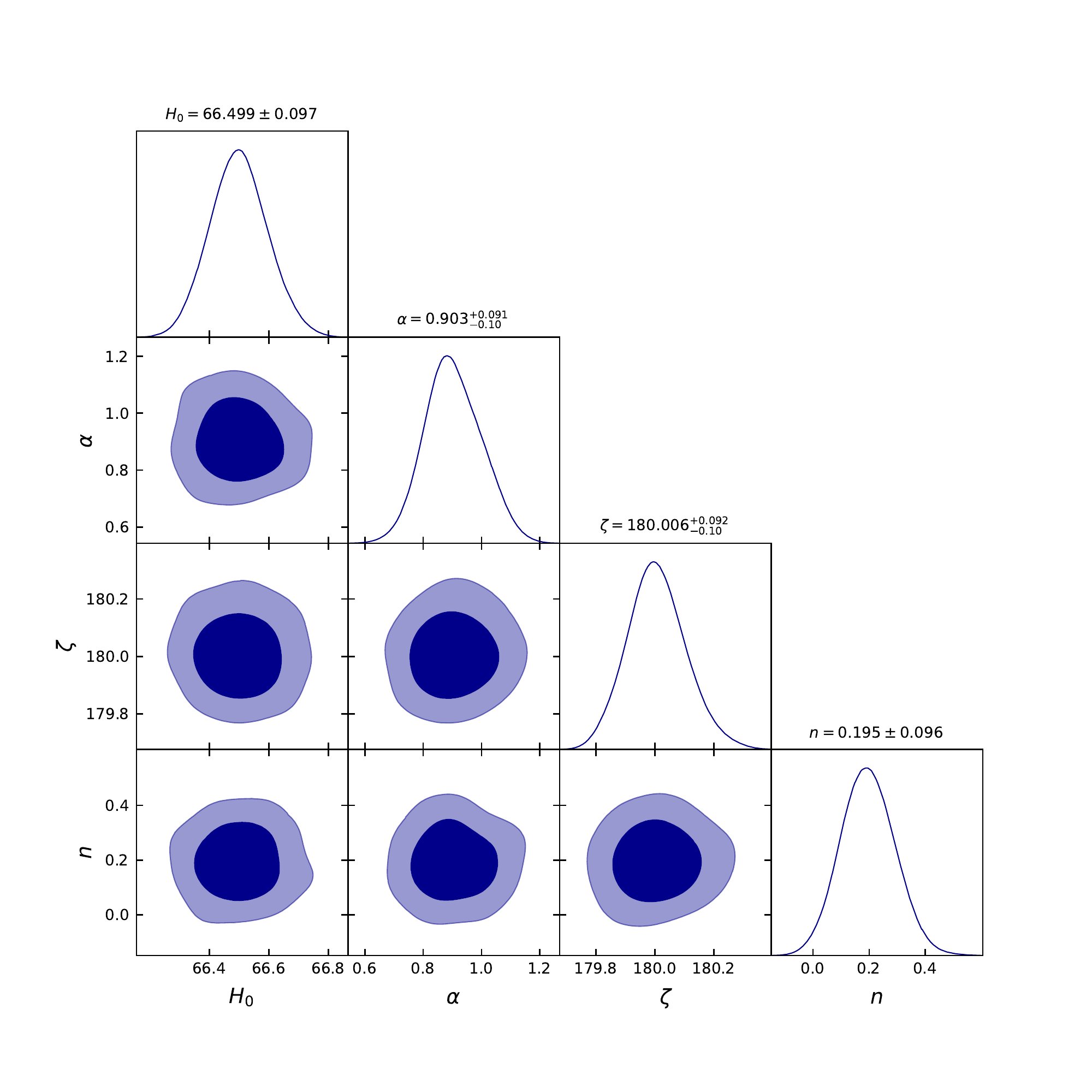}
\caption{The $1-\sigma$ and $2-\sigma$ contour plot for the model parameters using the H(z) data set.}\label{f1}
\end{figure}
\end{center}
\end{widetext}

\justify The obtained best fit values are  $\alpha=0.903^{+0.091}_{-0.10}$,  $\zeta=180.006^{+0.092}_{-0.10}$, $n=0.195 \pm 0.096$, and $H_0=66.499 \pm 0.097 $. 

\subsubsection{Pantheon datasets}
\justify
In this work, we incorporate 1048 points of Pantheon supernovae samples with corresponding distance moduli $\mu^{obs}$ in the redshift range $z \in [0.01, 2.3]$ \cite{Scolnic/2018}. The $\chi^2$ function corresponding to Pantheon data sets is given as
\begin{equation}\label{4b}
\chi^2_{SN}=\sum_{i,j=1}^{1048}\bigtriangledown\mu_{i}\left(C^{-1}_{SN}\right)_{ij}\bigtriangledown\mu_{j},
\end{equation}
Here $C_{SN}$ denotes the covariance matrix \cite{Scolnic/2018}, and
\begin{align*}\label{4c}
\quad \bigtriangledown\mu_{i}=\mu^{th}(z_i,\theta)-\mu_i^{obs}.
\end{align*} 
is the difference between the value of distance modulus obtained from the cosmological observations and its theoretically predicted values estimated from the given model with parameter space $\theta$. One can estimate the distance modulus by using the relation $\mu=m_B-M_B$, where $m_B$ and $M_B$ are the observed apparent magnitude and the absolute magnitude at a given redshift (Retrieving the nuisance parameter following the recent approach called BEAMS with Bias Correction (BBC) \cite{BMS}). Moreover, its theoretical value can be calculated as
\begin{equation}\label{4d}
\mu(z)= 5log_{10} \left[ \frac{D_{L}(z)}{1 Mpc}  \right]+25, 
\end{equation}
where 
\begin{equation}\label{4e}
D_{L}(z)= c(1+z) \int_{0}^{z} \frac{ dx}{H(x,\theta)}.
\end{equation}
\justify The contour plot for the model parameters $\alpha$, $\zeta$, $n$, and $H_0$ corresponding to Pantheon data sets at $1-\sigma$ and $2-\sigma$ confidence interval is presented in Fig \ref{f2}.

\begin{widetext}
\begin{center}
\begin{figure}[H]
\includegraphics[scale=0.8]{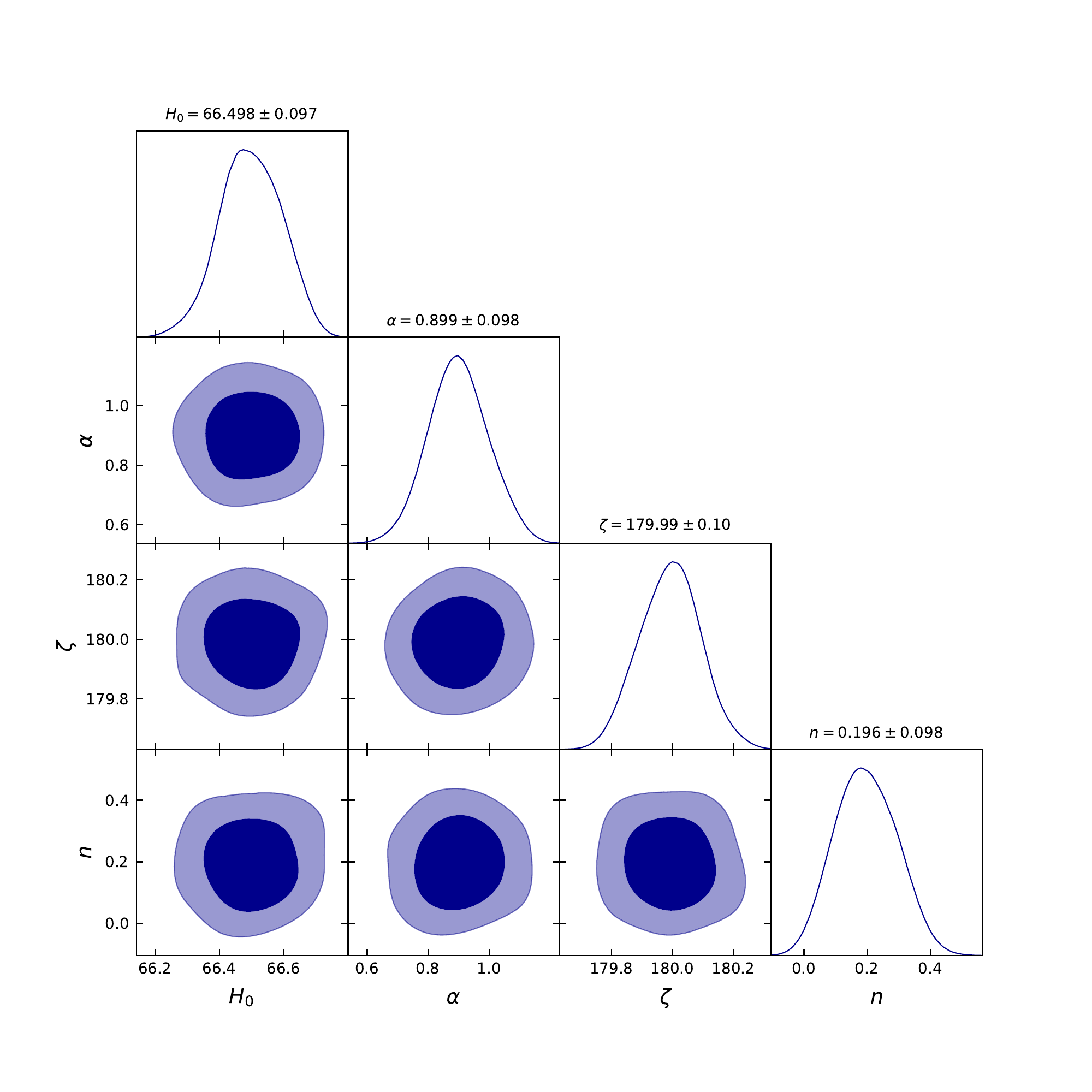}
\caption{The $1-\sigma$ and $2-\sigma$ contour plot for the model parameters using the Pantheon data set.}\label{f2}
\end{figure}
\end{center}
\end{widetext}

\justify The obtained best fit values are  $\alpha=0.899 \pm 0.098$,  $\zeta=179.99 \pm 0.10$, $n=0.196 \pm 0.098$, and $H_0=66.498 \pm 0.097 $. 

\subsubsection{H(z)+Pantheon datasets}

\justify The $\chi^{2}$ function to obtain the mean value of the model parameters corresponding to the combine H(z)+Pantheon data sets reads as

\begin{equation}
\chi^{2}_{total}= \chi^{2}_H + \chi^{2}_{SN} .
\end{equation}

\justify The contour plot for the model parameters $\alpha$, $\zeta$, $n$, and $H_0$ corresponding to H(z)+Pantheon data sets at $1-\sigma$ and $2-\sigma$ confidence interval is presented in Fig \ref{f3}.

\begin{widetext}
\begin{center}
\begin{figure}[H]
\includegraphics[scale=0.8]{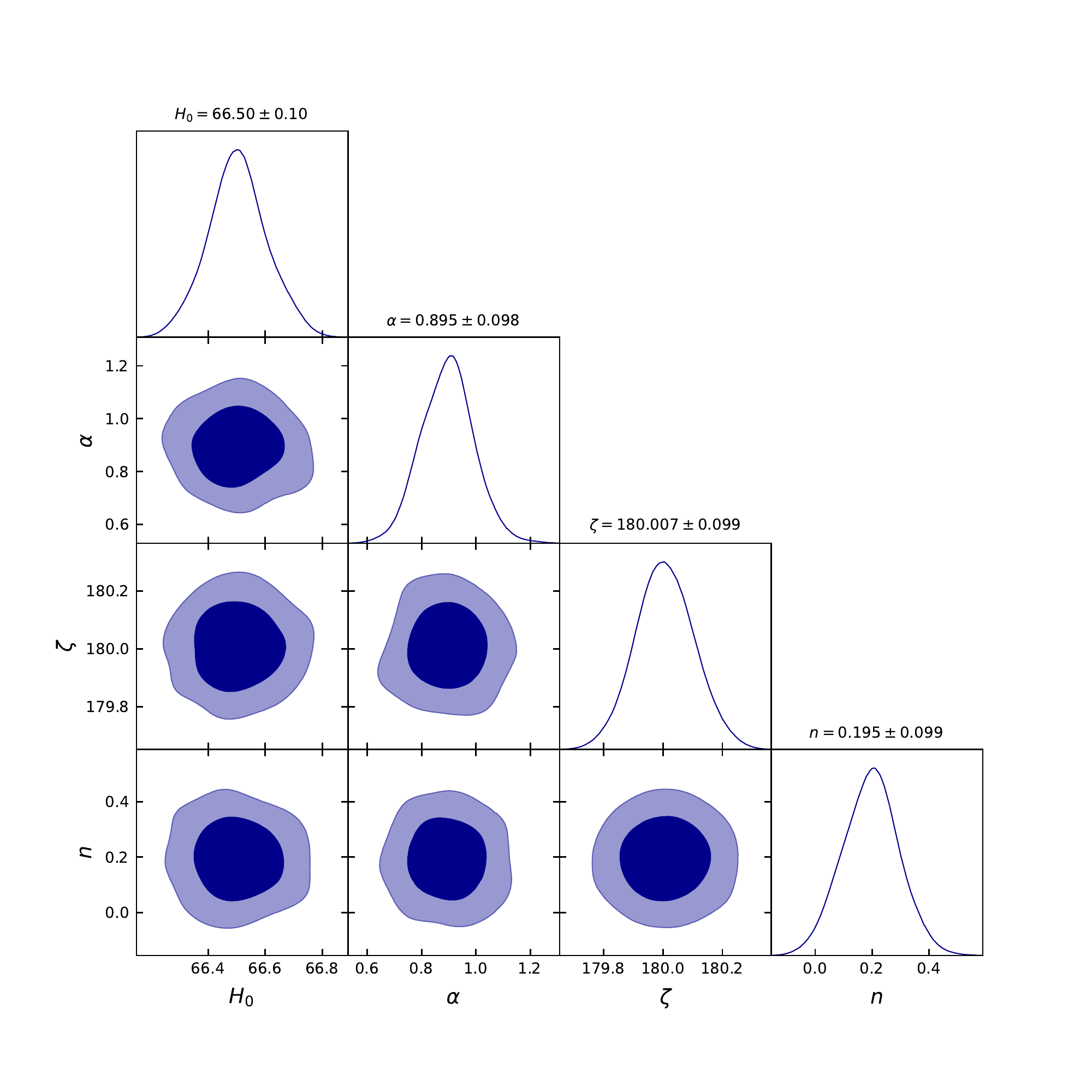}
\caption{The $1-\sigma$ and $2-\sigma$ contour plot for the model parameters using the H(z)+Pantheon data set.}\label{f3}
\end{figure}
\end{center}
\end{widetext}

\justify The obtained best fit values are  $\alpha=0.895 \pm 0.098$,  $\zeta=180.007 \pm 0.099$, $n=0.195 \pm 0.099$, and $H_0=66.50 \pm 0.10 $. 

\section{Behavior of Physical Parameters }\label{sec5}
\justify In this section, we are going to present the behavior of physical parameters that describes the evolution phase of the universe, corresponding to the extracted values of the model parameters from different observational data sets.

\begin{figure}[H]
\includegraphics[scale=0.5]{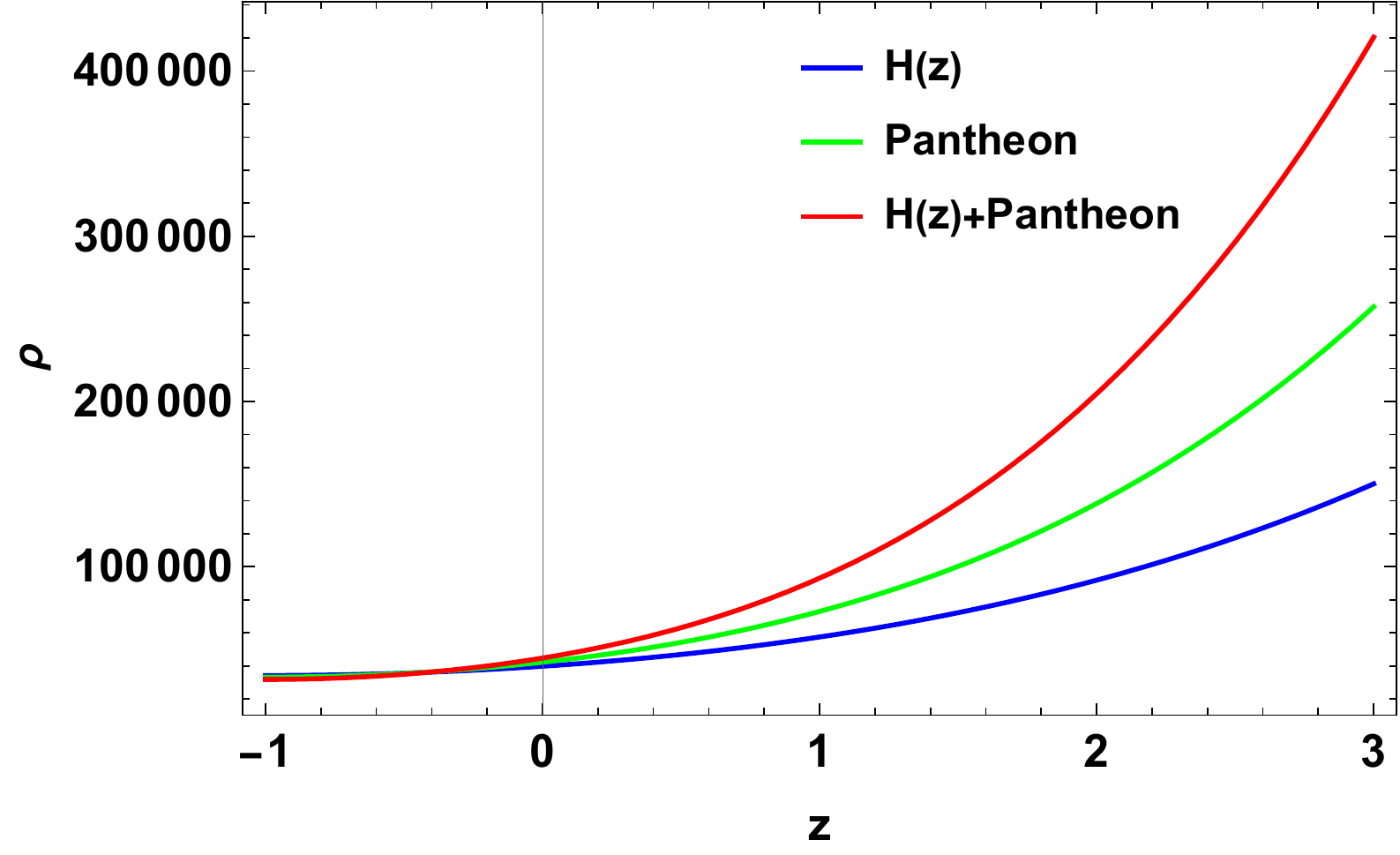}
\caption{Profile of the matter-energy density vs cosmic redshift $z$.}\label{f4}
\end{figure}

\begin{figure}[H]
\includegraphics[scale=0.51]{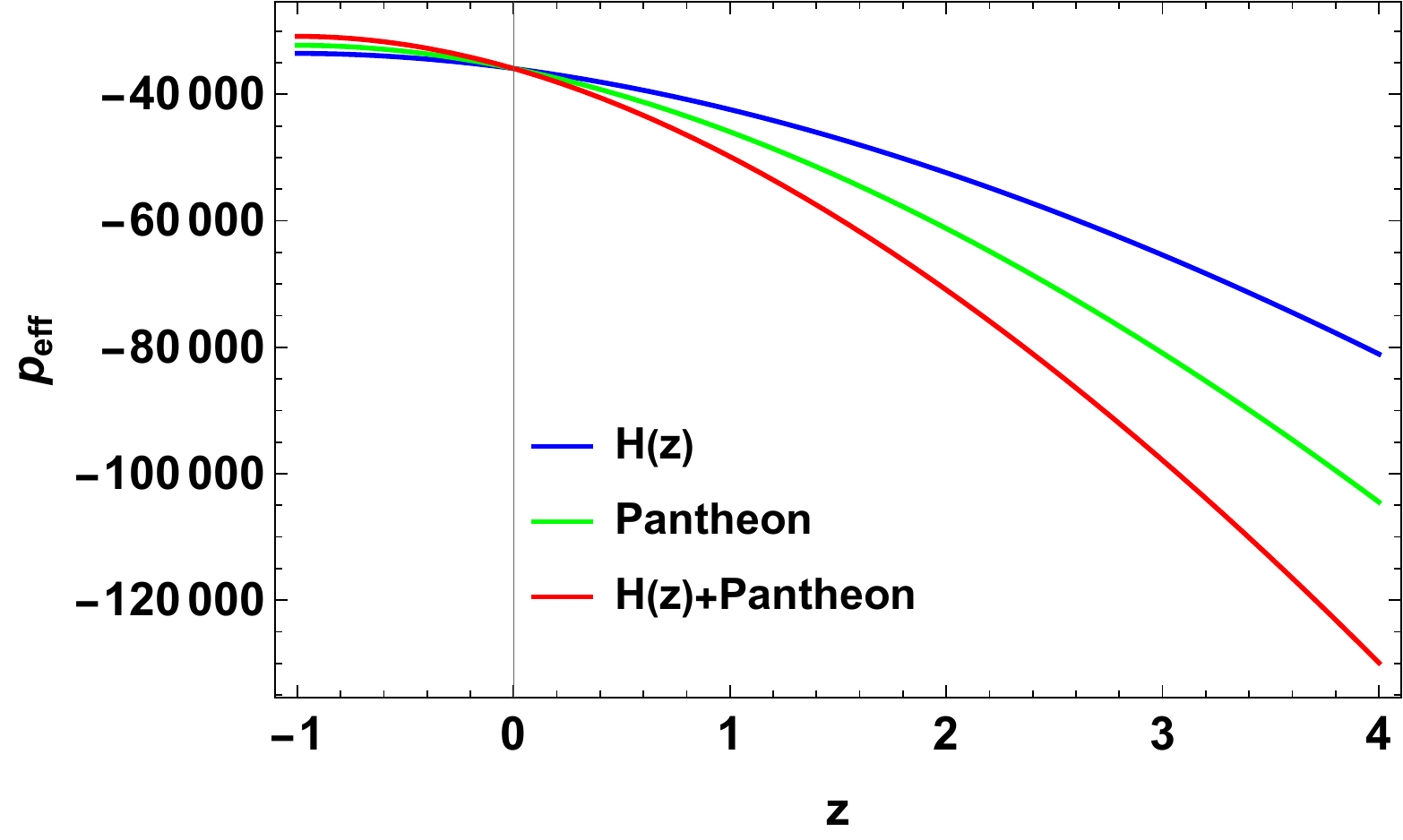}
\caption{Profile of the effective pressure vs cosmic redshift $z$.}\label{f5}
\end{figure}

\begin{figure}[H]
\includegraphics[scale=0.49]{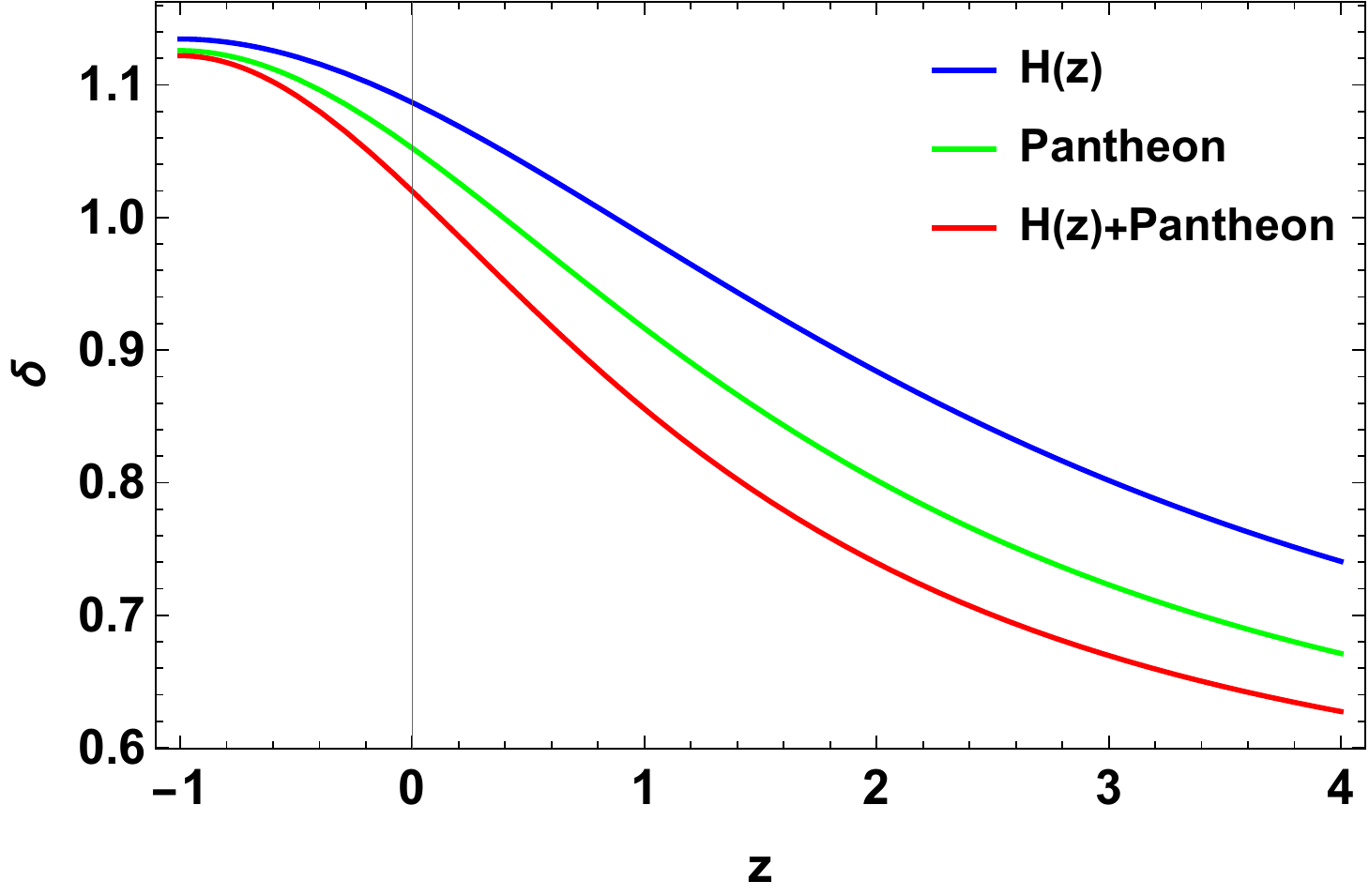}
\caption{Profile of the skewness parameter vs cosmic redshift $z$.}\label{f6}
\end{figure}

\begin{figure}[H]
\includegraphics[scale=0.47]{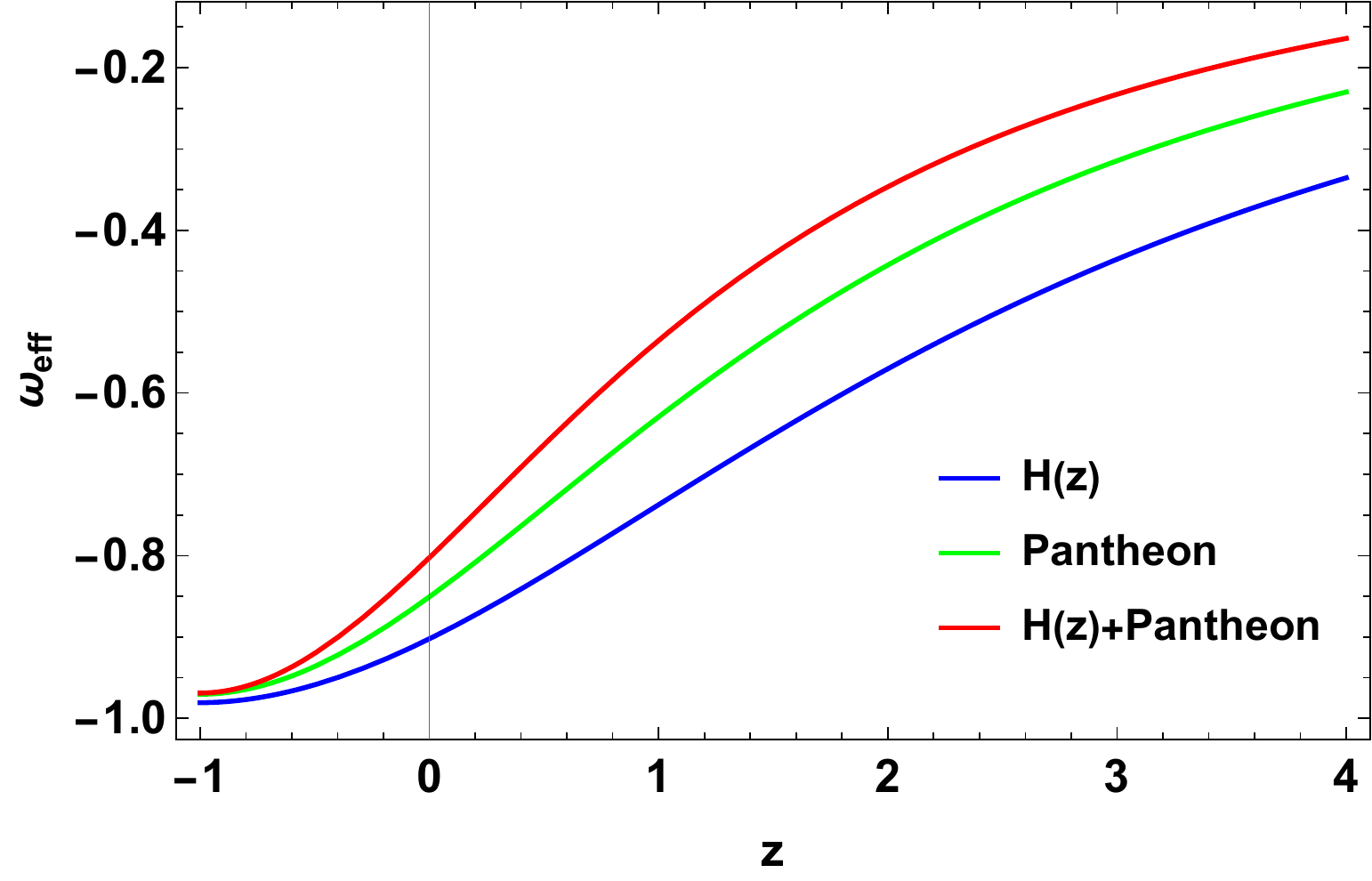}
\caption{Profile of the effective EoS vs cosmic redshift $z$.}\label{f7}
\end{figure}

\begin{figure}[H]
\includegraphics[scale=0.598]{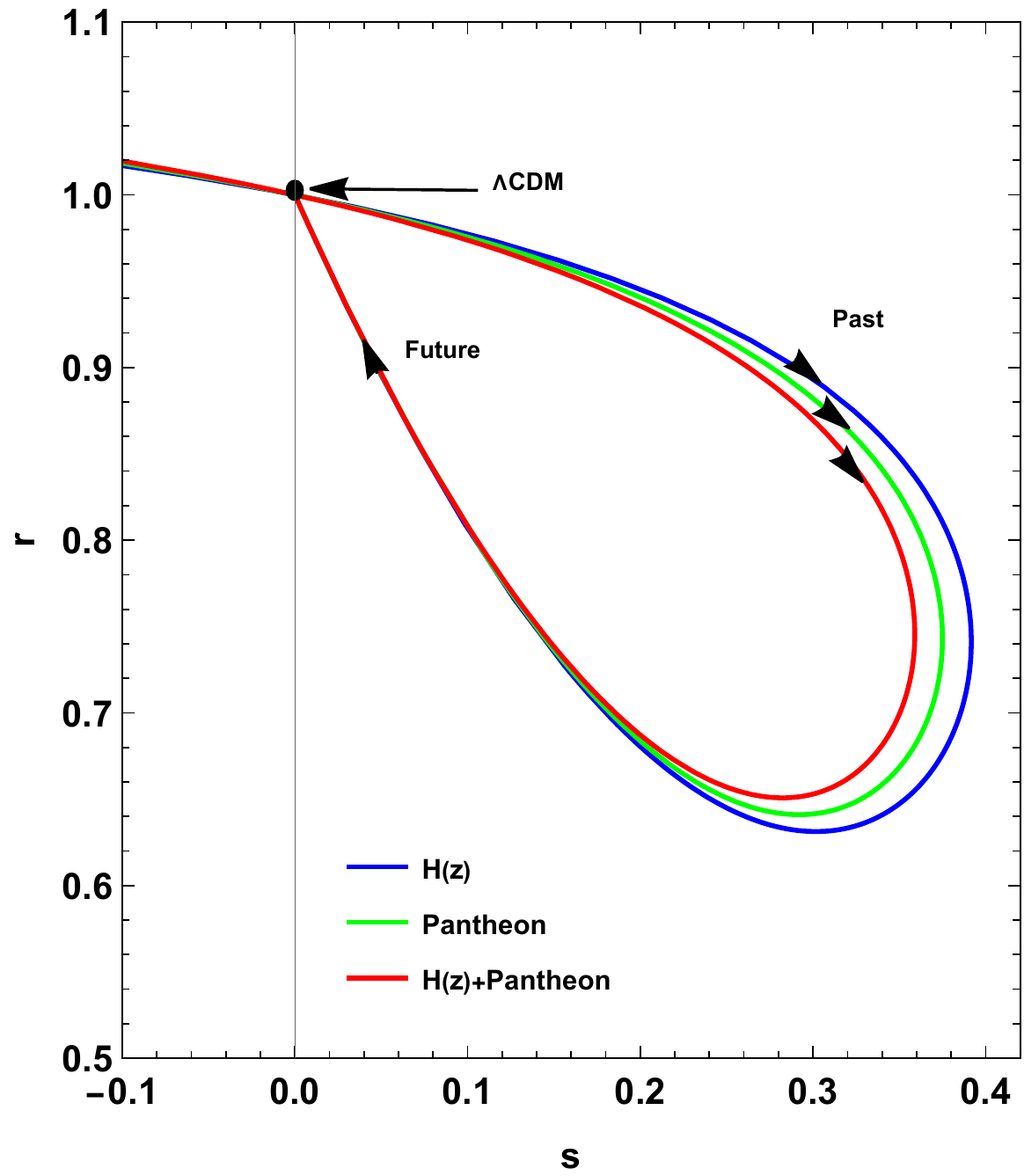}
\caption{Profile of evolutionary trajectories of the given model in the $r-s$ plane.}\label{f8}
\end{figure}

\justify The evolution profile of the matter-energy density and the effective pressure is presented in Fig \ref{f4} and \ref{f5}, respectively. We observe that the energy density indicates expected positive behavior, whereas pressure with viscosity coefficient shows negative behavior on the entire domain. Therefore the presence of the viscosity coefficient in the cosmic content contributes to the universe's expansion. The profile of the skewness parameter presented in Fig \ref{f6} shows the anisotropic nature of the spacetime during the entire evolution phase of the universe. Further, the effective EoS parameter presented in the Fig \ref{f7} favors the accelerating phase of the universe's expansion. It's present value corresponding to H(z), Pantheon, and combine H(z)+Pantheon data sets are $\omega_0 \approx -0.9 $, $\omega_0 \approx -0.85 $, and $\omega_0 \approx -0.8 $ respectively. 

The statefinder diagnostic proposed by V. Sahni \cite{V.S.} can geometrically differentiate different models of dark energy through statefinder parameters which are defined as 

\begin{equation}\label{5a}
 r=\frac{\dddot{a}}{aH^3} ,
\end{equation}
and
\begin{equation}\label{5b}
s=\frac{(r-1)}{3(q-\frac{1}{2})}.
\end{equation}

\justify The value $(r>1, s<0)$ represents phantom scenario whereas $(r<1, s>0)$ denotes the quintessence type dark energy, and the value $(r=1, s=0)$ mimics the standard $\Lambda$CDM model. The behavior of our $f(R,L_m)$ model in the presence of viscous fluid in the $r-s$ plane is presented in Fig \ref{f8}. The present values of statefinder parameters corresponding to H(z), Pantheon, and combine H(z)+Pantheon data sets are $(r,s)=(0.87,0.06)$, $(r,s)=(0.82,0.09) $, and $(r,s)=(0.77,0.12)$ respectively \cite{GBL}. We found that our cosmological $f(R,L_m)$ model lies in the quintessence region, and it behaves like the de-Sitter universe in the far future.

\section{Energy Conditions}\label{sec6}
\justify In this section, we investigate different energy conditions in order to test the consistency of the obtained solution of our model. The energy conditions are relations imposed on the energy-momentum tensor in order to satisfy positive energy. The energy conditions are obtained from the well-known Raychaudhuri equation and are written as \cite{EC} \\

\begin{itemize}
\item \textbf{Null energy condition (NEC) :} $\rho_{eff}+p_{eff}\geq 0$;  
\item \textbf{Weak energy condition (WEC) :} $\rho_{eff} \geq 0$ and  $\rho_{eff}+p_{eff}\geq 0$; 
\item \textbf{Dominant energy condition (DEC) :} $\rho_{eff} \pm p_{eff}\geq 0$; 
\item \textbf{Strong energy condition (SEC) :} $\rho_{eff}+ 3p_{eff}\geq 0$,
\end{itemize}
 with $\rho_{eff}$ being the effective energy density.

\begin{figure}[H]
\includegraphics[scale=0.53]{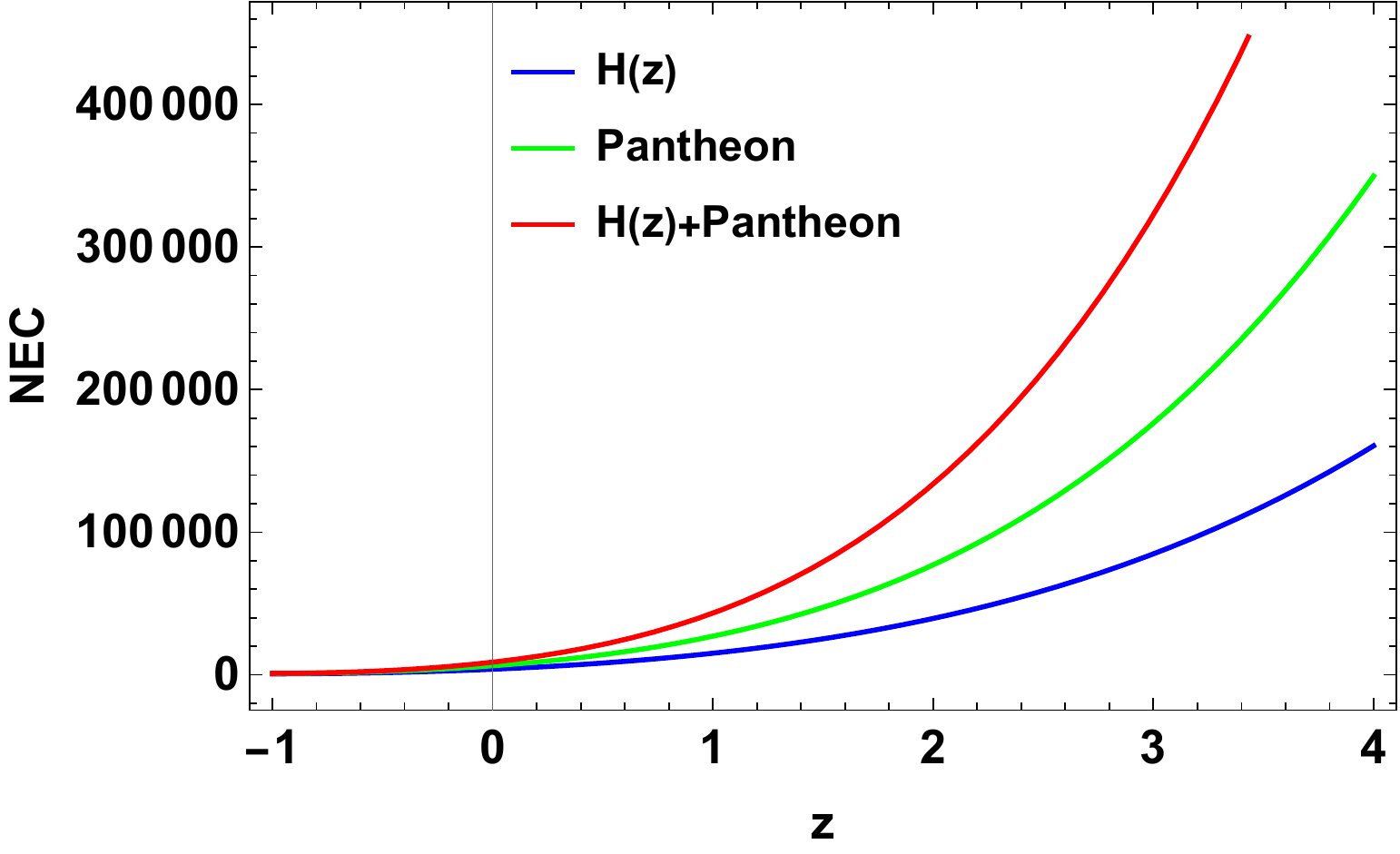}
\caption{Profile of the null energy condition vs cosmic redshift $z$.}\label{f9}
\end{figure}

\begin{figure}[H]
\includegraphics[scale=0.52]{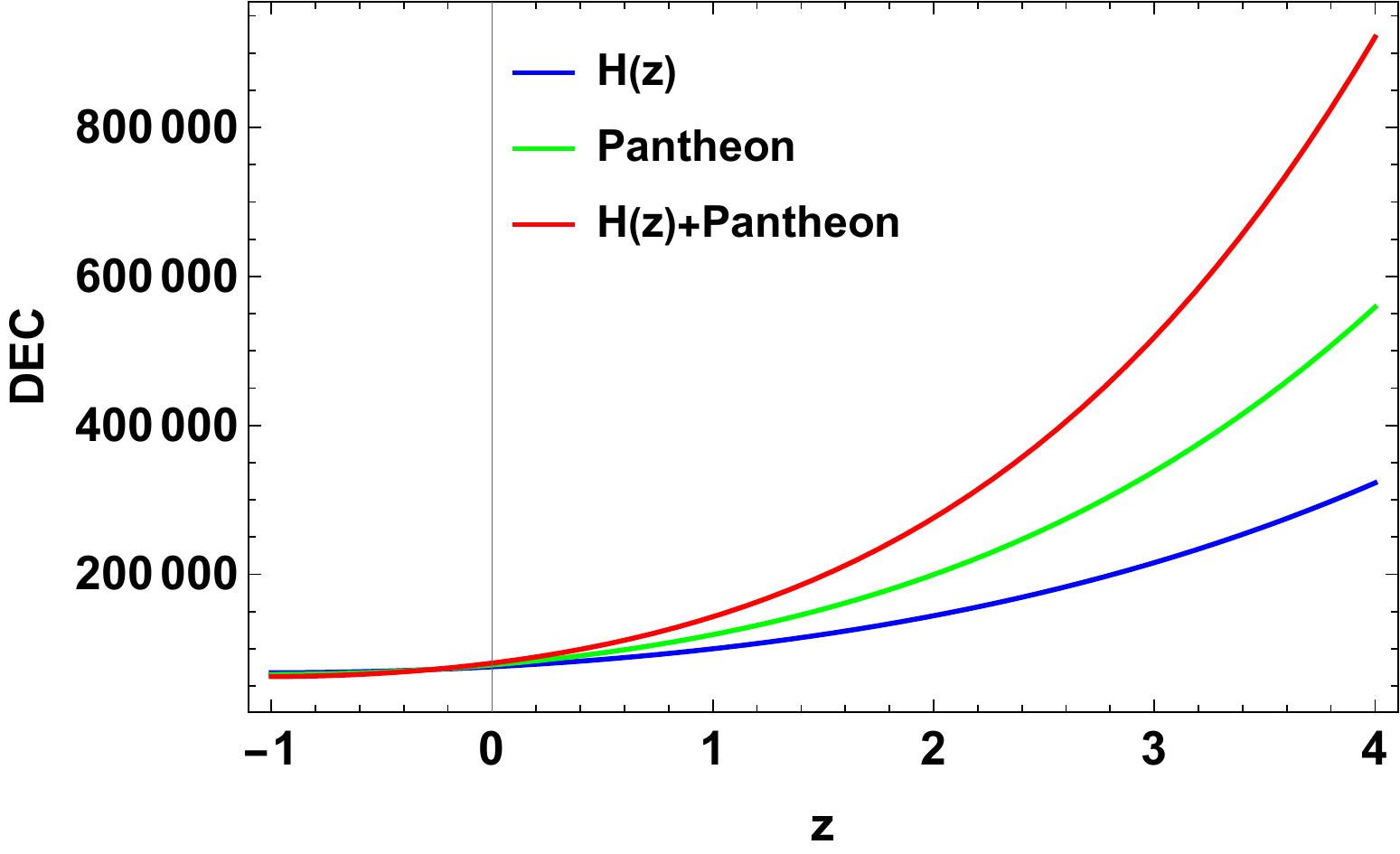}
\caption{Profile of the dominant energy condition vs cosmic redshift $z$.}\label{f10}
\end{figure}

\begin{figure}[H]
\includegraphics[scale=0.53]{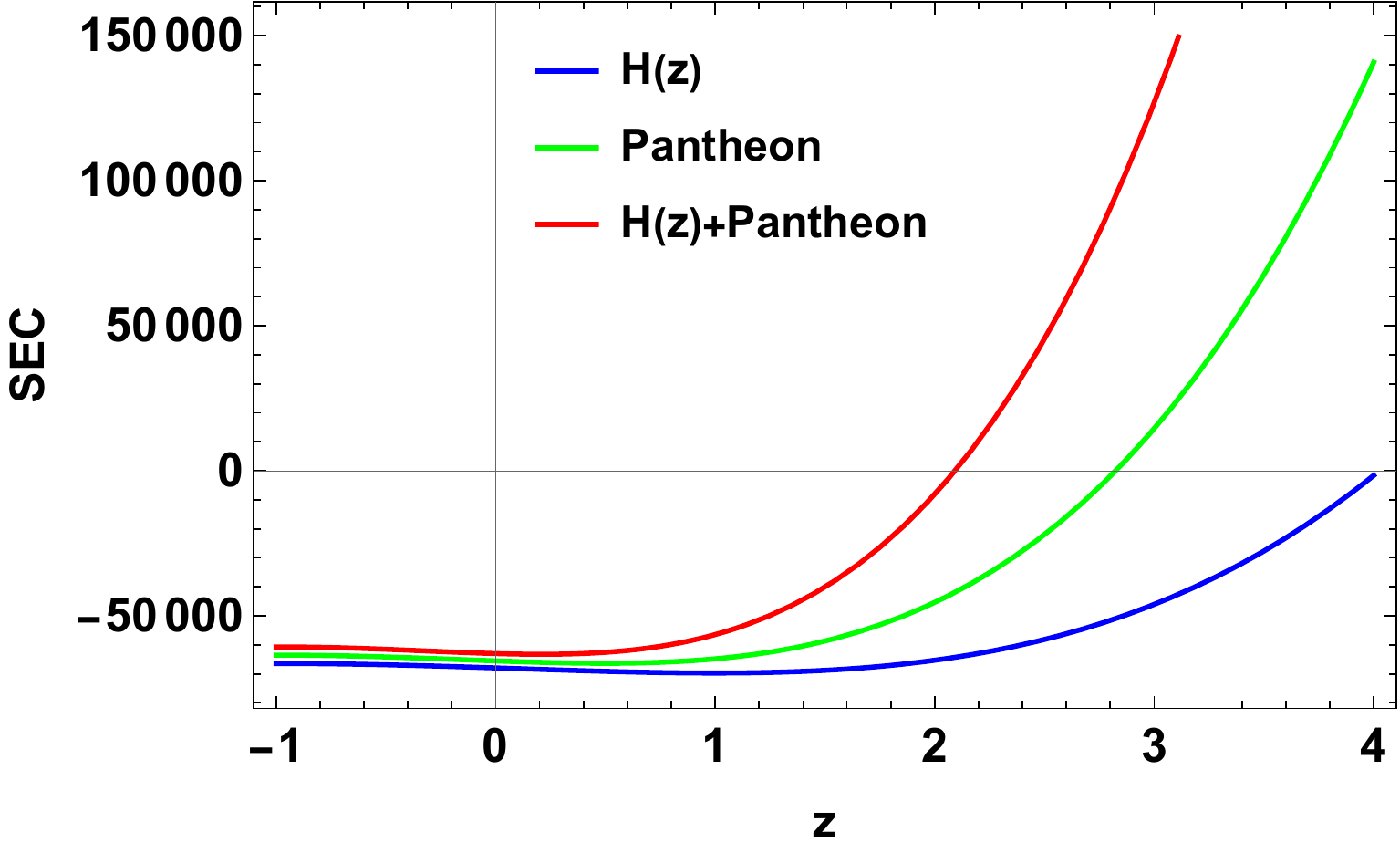}
\caption{Profile of the strong energy condition vs cosmic redshift $z$.}\label{f11}
\end{figure}

\justify The profile of the NEC and the DEC with respect to cosmic redshift is presented in Fig \ref{f9} and \ref{f10}, respectively. Since WEC is the combination of NEC and the positive energy density, we found that corresponding to the extracted values of the model parameters from different observational data sets, NEC, DEC, and WEC shows positive behavior, whereas SEC presented in Fig \ref{f11} favors a transition from positive to negative behavior in the recent past. Thus, the violation of SEC strongly supports the recently observed acceleration with the transition from decelerated to the accelerated phase of the universe's expansion in the recent past.

\section{Conclusion}\label{sec7}

\justify From the hydrodynamics perspective, the assumption of viscous effects in the cosmic fluid is quite natural since the perfect fluid is, after all, an abstraction. In this work, we investigated the significance of viscosity coefficients to describe the observed cosmic acceleration by taking into account a cosmological $f(R,L_m)$ model with an anisotropic background. We considered $f(R,L_m)=\frac{R}{2}+L_m^\alpha $, where $\alpha$ is an arbitrary parameter, with the effective equation of state $\bar{p}=p-3 \zeta H$ that is the Einstein case value with proportionality constant $\zeta$ used in the Einstein theory \cite{IB-0}. In section \ref{sec3}, we obtained the analytical expression of the Hubble parameter H(z) by incorporating the assumed $f(R,L_m)$ viscous fluid model. Further, in section \ref{sec4}, we analyzed the viability of the assumed $f(R,L_m)$ model by incorporating the observational data sets, specifically, H(z) data sets and the Pantheon supernovae data sets. The obtained best fit values are  $\alpha=0.903^{+0.091}_{-0.10}$,  $\zeta=180.006^{+0.092}_{-0.10}$, $n=0.195 \pm 0.096$, and $H_0=66.499 \pm 0.097 $ for the H(z) datasets, $\alpha=0.899 \pm 0.098$,  $\zeta=179.99 \pm 0.10$, $n=0.196 \pm 0.098$, and $H_0=66.498 \pm 0.097 $ for the Pantheon datasets, and $\alpha=0.895 \pm 0.098$,  $\zeta=180.007 \pm 0.099$, $n=0.195 \pm 0.099$, and $H_0=66.50 \pm 0.10 $ \cite{HHH} for the combined H(z)+Pantheon datsets. In section \ref{sec5}, behavior of some cosmological parameters have been presented. We observed that the energy density presented in Fig \ref{f4} decreases with the expansion of the universe, whereas the bulk viscous pressure presented in Fig \ref{f5} indicates the negative behavior. The skewness parameter presented in Fig \ref{f6} favors the anisotropic type evolution of the universe during the entire time regime. Further, the effective EoS parameter in Fig \ref{f7} shows the accelerating nature of the cosmic expansion with the present values $\omega_0 \approx -0.9 $, $\omega_0 \approx -0.85 $, and $\omega_0 \approx -0.8 $ corresponding to H(z), Pantheon, and the combined H(z)+Pantheon data sets respectively. Finally, from Fig \ref{f8} we found that our cosmological $f(R,L_m)$ model exhibits quintessence behavior, and it favors the de-Sitter type expansion in the far future. In section \ref{sec6}, energy conditions have been analyzed to interpret the viability of the obtained solution. All energy conditions except SEC exhibit positive behavior (see Fig \ref{f9} and \ref{f10}), whereas the violation of SEC presented in Fig \ref{f11} strongly supports the accelerating nature of cosmic expansion with the transition from decelerated to accelerated era.

\section*{Data Availability Statement}

There are no new data associated with this article.

\section*{Acknowledgments} \label{sec10}
R.S. acknowledges UGC, New Delhi, India for providing Senior Research Fellowship (UGC-Ref. No.: 191620096030). L.V.J. acknowledges UGC, Govt. of India, New Delhi, for awarding JRF (NTA Ref. No.: 191620024300). PKS acknowledges Science and Engineering Research Board, Department of Science and Technology, Government of India for financial support to carry out Research project No.: CRG/2022/001847 and IUCAA, Pune, India for providing support through the visiting Associateship program. We are very much grateful to the honourable referees and the editor for the illuminating suggestions that have significantly improved our research quality and presentation.

\end{document}